# Tensor-monopole-induced topological boundary effects in four-dimensional acoustic metamaterials

Qingyang Mo[1*], Shanjun Liang[2*], Cuicui Lu[1,3§], Jie Zhu[4‡], Shuang Zhang[1,5,6,7†]

[1] New Cornerstone Science Laboratory, Department of Physics, University of Hong Kong; 999077, Hong Kong, China
[2] School of Professional Education and Executive Development, The Hong Kong Polytechnic University, Hong Kong, China
[3] Key Laboratory of Advanced Optoelectronic Quantum Architecture and Measurements of Ministry of Education, Beijing Key Laboratory of Nanophotonics and Ultrafine Optoelectronic Systems, Center for Interdisciplinary Science of Optical Quantum and NEMS Integration, School of Physics, School of Physics, Beijing Institute of Technology, Beijing 100081, China
[3] School of Professional Education and Executive Development, The Hong Kong Polytechnic University, Hong Kong, China
[4] Institute of Acoustics, School of Physics Science and Engineering, Tongji University, 200092 Shanghai, China
[5] Department of Electrical & Electronic Engineering, University of Hong Kong; 999077, Hong Kong, China
[6] Quantum Science Center of Guangdong-Hong Kong-Macao Great Bay Area, 3 Binlang Road, Shenzhen, China
[7] Materials Innovation Institute for Life Sciences and Energy (MILES), HKU-SIRI, Shenzhen, P.R. China

Gauge field theory provides the mathematical and conceptual framework to describe and understand topological singularities such as Weyl points and magnetic monopoles. While singularities associated with vector electromagnetic gauge fields have been well-studied, those of higher-form tensor gauge fields, like the four-dimensional (4D) tensor monopoles predicted by string theory, have remained largely theoretical or limited to experimental demonstration in pure synthetic dimensions, thereby not allowing investigations of the associated boundary effects. Here, we present a 4D system with tensor monopoles using engineered acoustic metamaterials. Our momentum space combines three real momentum dimensions and a geometric parameter as the fourth. By varying this fourth momentum, we experimentally reveal two distinct topological surface states in 3D subsystems: Fermi-arc surface states in a gapless subsystem and Dirac-cone surface states in a gapped subsystem. Our work introduces a novel platform for exploring new topological structures associated with tensor gauge field and topological phenomena in higher dimensions.

Gauge theory provides a powerful framework for describing the dynamics of elementary particles and gravitation [1–3]. In electromagnetism, vector gauge fields, such as electromagnetic gauge potentials, predict magnetic monopoles in odd dimensions, including U(1) Abelian Dirac monopoles in three dimensions and U(2) non-Abelian Yang monopoles in five dimensions [4–6]. While magnetic monopoles remain elusive in high-energy physics, their momentum-space counterparts have been identified in natural materials [7–10], ultracold matter [11–13], and artificial structures [14–20]. A prominent example is the Weyl point, a linear band crossing in 3D crystals that acts as a monopole source of Berry curvature. Weyl semimetals hosting pairs of Weyl points exhibit exotic topological phenomena, such as Fermi-arc surface states [21–24], chiral zero modes [25–27], and quantum oscillations [28].

Extending vector gauge fields to higher forms, known as tensor Kalb-Ramond gauge fields [29,30], has opened new avenues for exploring higher-dimensional topological phases [31–37]. In this context, the simplest generalization of magnetic monopoles is the "tensor monopole", a point defect of 2-form tensor gauge fields in four dimensions characterized by the Dixmier-Douady (DD) invariant [38,39]. Higher-form gauge fields have been proposed as tools for studying exotic topological phases, including 3D chiral topological insulators [40–43], higher-order topology [44], and fracton matter [45,46]. However, tensor monopoles and related tensor gauge fields have not been realized in real momentum dimensions, which are necessary for the study of their topological boundary states.

In recent years, acoustic systems have emerged as a versatile platform for exploring topological phenomena through the manipulation of gauge fields. By designing acoustic metamaterials with tailored lattice structures or synthetic dimensions, researchers have successfully realized a variety of topological effects, such as Landau levels [47–49], Chern insulators [50,51], and Weyl semimetals [16]. These studies demonstrate the unique capability of acoustic systems to emulate complex gauge fields and explore topological phases that are difficult to access in natural materials.

In this work, we demonstrate 4D acoustic metamaterials, constructed by three real dimensions and one synthetic momentum dimension, that host chiral symmetric tensor monopoles. In analogy to the Weyl semimetals, the boundary states of the tensor monopole semimetals (TMSMs) take the form of Dirac-cone arcs, connecting the projections of opposite tensor monopoles. We experimentally observe key features of boundary states of TMSMs in two selected 3D subspaces through judicious design of acoustic metamaterials, i.e., Fermi-arc surface states in a gapless subsystem and Dirac-cone surface states in a gapped subsystem. Our work provides a practical material platform for exploring topological phenomena related to tensor gauge field.

We start from the minimal Hamiltonian for tensor monopole residing in 4D parameter space [32,33]:

$$H_{TM} = \Phi_1 \lambda_4 + \Phi_2 \lambda_5 + \Phi_3 \lambda_7 + \Phi_4 \lambda_6$$
$$= \begin{pmatrix} 0 & 0 & \Phi_1 - i\Phi_2 \\ 0 & 0 & \Phi_4 - i\Phi_3 \\ \Phi_1 + i\Phi_2 & \Phi_4 + i\Phi_3 & 0 \end{pmatrix}, \quad (1)$$

where $\vec{\Phi} = (\Phi_1, \Phi_2, \Phi_3, \Phi_4)$ represent the 4D parameter space and $\lambda_i$ are $3 \times 3$ Gell-Mann matrices [52]. As shown in Fig. 1(a, b), a triply degenerated nodal point located at the origin $\vec{\Phi} = 0$ serves as a monopole source of 3-form Berry curvature $\mathcal{H}_{\mu\nu\lambda} = \partial_\mu B_{\nu\lambda} + \partial_\nu B_{\lambda\mu} + \partial_\lambda B_{\mu\nu}$, where $B_{\mu\nu}$ represents a two-form antisymmetric tensor gauge field [53]. The topological charge of tensor monopole is determined by the integral of the 3-form Berry curvature on a $S^3$ manifold enclosing this nodal point, known as the DD invariant,

$$Q_{DD} = \frac{1}{2\pi^2} \int_{S^3} \mathcal{H}_{\mu\nu\lambda} d\Phi^\mu \wedge d\Phi^\nu \wedge d\Phi^\lambda = 1. \quad (2)$$

It is worth noting that the tensor monopole is protected by chiral symmetry $\{H_{TM}, U\} = 0$, where $U = diag(1,1,-1)$, and serves as the AIII-class monopole in four dimensions predicted by the ten-fold way topological classification [54].

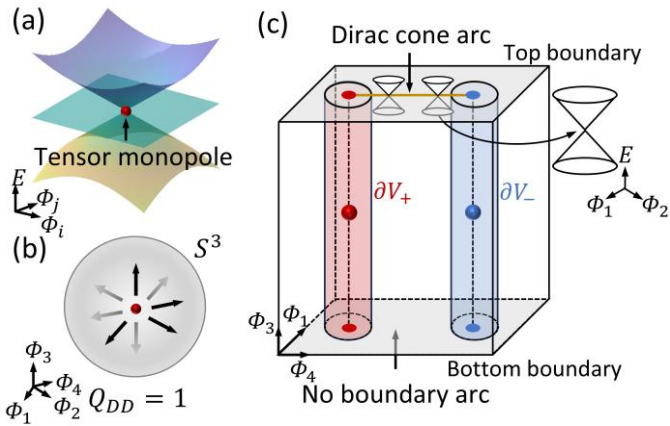

FIG. 1. (a) The dispersion spectrum for the tensor monopole system. (b) The distribution of 3-form Berry curvature in the 4D space and the DD invariant $Q_{DD} = 1$. (c) The boundary states of TMSMs behave as a Dirac-cone arc (the yellow line in the top boundary of $\phi_3$ direction) and cross all possible $\partial V_\pm$. Red or blue dot denotes tensor monopoles with $Q_{DD} = \pm 1$. $\phi_2$ is fixed to 0.

Interestingly, the DD invariant of tensor monopoles leads to a Dirac-cone arc on the 3D boundary of TMSMs. As shown in Fig. 1(c), the 3D manifold $\partial V_\pm$ surrounding the tensor monopole (red or blue dot) corresponds to a 3D gapped system with DD invariant $Q_{DD} = \pm 1$. Such a 3D gapped system can be considered a Chiral topological insulator (CTI), hosting surface states in the form of Dirac cones. Interestingly, by setting an open boundary condition for this CTI in one direction, e.g. the $\Phi_3$ direction, a Dirac cone appears on only one surface (e.g. the top surface) [53]. Furthermore, the manifold $\partial V_\pm$ can be freely deformed without altering its topological properties as long as it contains a single tensor monopole. Therefore, a 1D arc of 2D Dirac cones appears on one boundary that crosses any possible $\partial V_\pm$, shown as the yellow arc in the $\Phi_2 = 0$ subspace in Fig. 1(c).

Next, we propose an acoustic metamaterial, with a coupling term serving as the fourth momentum component, as a realistic platform for the investigation of exotic topological boundary phenomena in TMSMs. Fig. 2(a, b) shows the unit cell of the deformed hexagonal acoustic lattice and the corresponding tight-binding model (TBM) [53]. The acoustic cavities (yellow cubes) correspond to three sites (labeled as "A", "B", "C") in the TBM and support the dipolar resonance mode around 8550 Hz. The gray tunnels connecting A and C cavities correspond to nearest-neighbor couplings $t_1$ in the $x - y$ plane, while blue (red) tunnels connecting B and C cavities correspond to positive intra- (negative inter-) couplings $t_{2(3)}$ in the $z$ direction. The fourth synthetic momentum vector $k_w$ is introduced by modulating z-directional couplings as functions of $t_2 = t_0(1 - \xi \cos k_w)$ and $t_3 = -t_0(1 + \xi \cos k_w)$, where $\xi$ is a dimensionless coefficient. Fig. 2(c) shows the theoretic dispersion spectrum in the 4D Brillouin zone (BZ) with $t_0 = 270\ Hz$, $t_1 = 450\ Hz$, and $\xi = 0.6$. It reveals two pairs of tensor monopoles (red and blue dots) located at $K$ and $K'$ valleys with $k_w = \pm \pi/2$ [53]. To realize tensor monopoles in acoustic metamaterials, we adjust the cross-sectional areas of intra-cell (blue) and inter-cell (red) tunnels as functions of $s_1 = s_0(1 - \xi \cos k_w)$ and $s_2 = s_0(1 + \xi \cos k_w)$, where $s_0 = 5mm \times 5mm$ and $\xi = 0.6$, to modulate the coupling strength between acoustic cavities in the z direction. The simulated dispersion spectrum is plotted in Fig. 2(d), which agrees well with the theoretical result in Fig. 2(c).

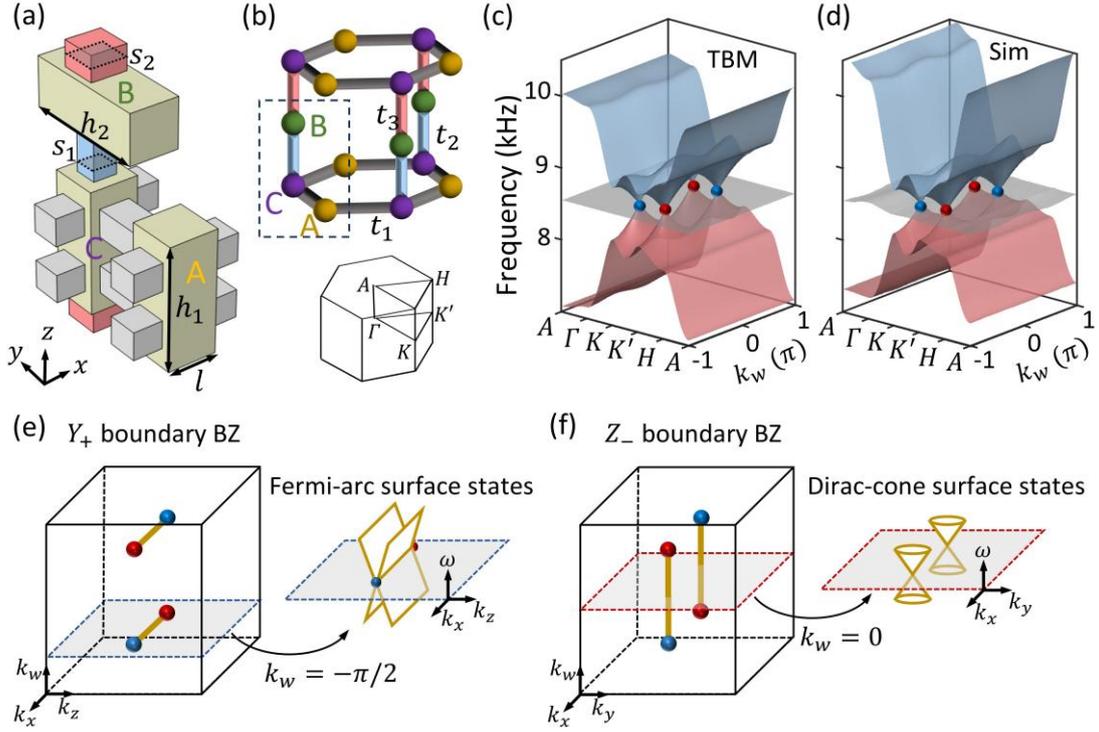

FIG. 2. (a, b) Acoustic unit cell (a) and corresponding tight-binding model (b) for the tensor monopole semimetals (TMSMs). The inset shows high-symmetry points in the real 3D first Brillouin zone (BZ). The structure parameters are $h_1 = 20mm$, $h_2 = 20.7mm$, $l = 8mm$. The lattice constants $a_x = a_y = 17.6mm$, $a_z = 44.4mm$. The lattice vectors are $\vec{a}_1 = (a_x, a_y, 0)$, $\vec{a}_2 = (a_x, -a_y, 0)$, and $\vec{a}_3 = (0, 0, a_z)$. (c, d) The theoretic and simulated dispersion spectrum for TMSMs, respectively. Red and blue dots denote positive and negative tensor monopoles. (d) Left: the distribution of Dirac cone arcs (yellow lines) in the $Y_+$ boundary BZ. Right: The dispersion spectrum for the Fermi-arc surface states on the $Y_+$ surface of gapless subsystem ($k_w = -\pi/2$). (e) Left: the distribution of Dirac cone arcs in the $Z_-$ boundary BZ. Right: The dispersion spectrum for the Dirac-cone surface states on the $Z_-$ surface of gapped subsystem ($k_w = 0$).

According to the bulk-boundary correspondence, there should exist topological boundary states in 3D boundary BZs. Note that Dirac cone arcs cannot exist on X or W boundaries, because in these two projecting axes the projections of two opposite tensor monopoles would coincide making the topological charges cancel out. As shown in Fig. 2(e, f), Dirac cone arcs (yellow lines) connect projections of opposite tensor monopoles along the $k_x$ direction on the positive Y ($Y_+$) boundary and along the $k_w$ direction on the negative Z ($Z_-$) boundary, respectively [53].

To capture the key features of boundary states, we select two different 3D subspaces (gapless and gapped subspaces) by carefully choosing the synthetic momentum vector $k_w$. In the 3D gapless subsystem ($k_w = -\pi/2$), the $Y_+$ surface corresponds to the gray plane containing Dirac cone arcs in left panel of Fig. 2(e), where the Fermi-arc surface states are formed by two intersecting planes of linear dispersion, as shown in right panel of Fig. 2(e). Interestingly, the tensor monopoles serve as dipole sources of conventional 2-form Berry curvature in the 3D subspace [53]. In contrast, in the 3D gapped subsystem ($k_w = 0$), the $Z_-$ surface, corresponding to the gray plane orthogonal to Dirac-cone arcs in left panel of Fig. 2(f), supports Dirac-cone surface states, as shown in right panel of Fig. 2(f). Section 8 of [53] presents simulated Fermi-arc and Dirac-cone surface states in the gapless and gapped subsystem, respectively.

We first experimentally demonstrate a gapless subsystem of TMSMs with fixed $k_w = -\pi/2$, and verify the existence of tensor monopoles and Fermi-arc surface states. We use one-step 3D printing to manufacture the acoustic sample with a fabrication precision of 0.1mm. The photo of the fabricated acoustic metamaterials is shown in Fig. 3(a). The inset shows the corresponding hollow acoustic unit with $s_1 = s_2 = s_0$, which is surrounded by a resin wall of 3 mm thickness. Fig. 3(b) shows the theoretic (red dashed line) and simulated (gray solid line) band dispersion between high-symmetry points, which indicates a pair of tensor monopoles located at $K$ and $K'$ valleys. First, we obtain the projected bulk bands on the $Y_-$ surface by carrying out acoustic measurements [53]. The measured iso-frequency surfaces for projected bulk states are plotted together with the numerical results (white dashed lines) in Fig. 3(c). When the frequency increases from 8250 Hz to 9050 Hz, the projected bulk bands shrink from two ellipse areas to two crossing points at $(k_x, k_z) = (\pm\frac{\pi}{3a_x}, 0)$ at the frequency of 8600 Hz, and then expand to ellipse areas again, giving direct evidence of the existence of tensor monopoles.

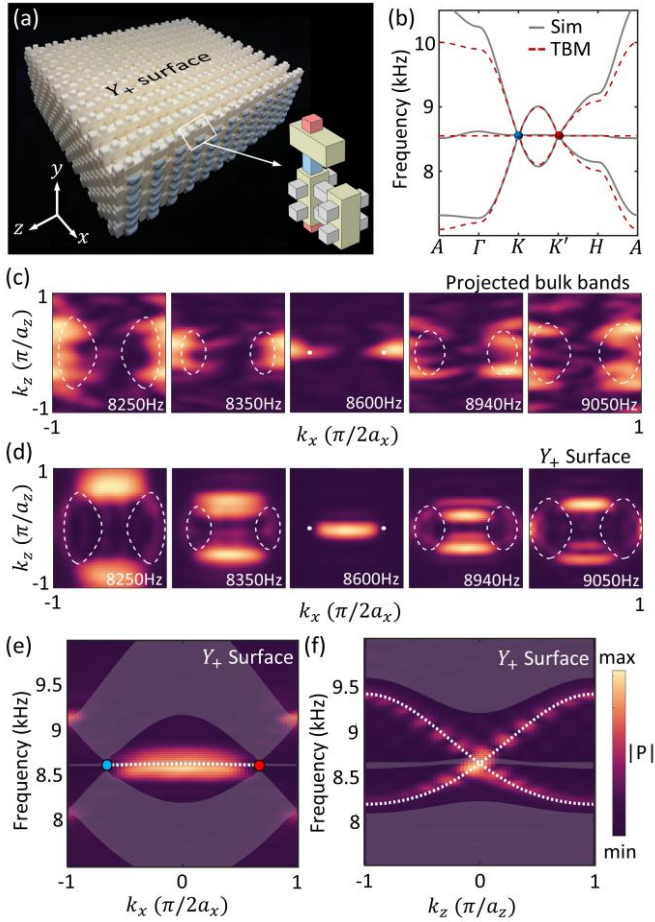

FIG. 3. (a) The photo of acoustic sample for the gapless subsystem, containing $12 \times 8 \times 12$ cells. Inset: the hollow acoustic unit with $s_1 = s_2 = s_0$. (b) The theoretic and simulated dispersion spectra along the $A - \Gamma - K - K' - H - A$ high-symmetry line. (c, d) Measured iso-frequency surfaces for projected bulk states (c) and Fermi-arc surface states (d) at the frequency of 8250 Hz, 8350 Hz, 8600 Hz, 8940 Hz, and 9050 Hz, respectively. White dashed lines and dots represent projected bulk bands and tensor monopoles from simulation. (e, f) Measured spectra of Fermi-arc surface states along the $k_x$ direction with $k_z = 0$ (e) and along the $k_z$ direction with $k_x = 0$ (f). Gray areas and white dashed lines denote simulated projected bulk and surface states, respectively.

We next experimentally study the gapped subsystem of TMSMs with fixed $k_w = 0$, and show the observation of Dirac-cone surface states. The photo of fabricated sample and the corresponding hollow acoustic unit surrounded by a resin wall of 3 mm thickness are shown in Fig. 4(a), where $s_1 = 0.4s_0$ and $s_2 = 1.6s_0$. By applying open boundary condition in the $z$ direction, we plot the projected dispersion obtained by TBM and simulation in Fig. 4(b, c), respectively. At the $K$ valley, the projected bulk bands (gray areas) are gapped out, while a Dirac one exists on the $Z_-$ surface. The conical dispersion of the surface states in the 2D reciprocal space is shown in Fig. 4(d), which exhibits linear dispersion in $k_x$ ($k_y$) momentum space. There exists another surface Dirac cone located at the $K'$ valley, which is not shown in this plot. The measured spectrum of projected bulk bands between high-symmetry points is shown in Fig. 4(e), which exhibit a band gap of nearly 700 Hz at the $K$ valley. Note that the acoustic measurements are only performed at the C sites, where the middle flat bulk bands are not excited or measured [53]. The measured dispersion on the $Z_-$ surface is shown in Fig. 4(f), which clearly shows a linear band crossing, i.e. a surface Dirac point, at the $K$ valley at the frequency around 8600 Hz. The measured iso-frequency surfaces at a number of frequencies ranging from 8200 Hz to 9000 Hz are shown in Fig. 4(g), which reveals a conical dispersion structure, confirming the existence of Dirac-cone surface states in the gapped subsystem of TMSMs.

To sum up, we have constructed tensor monopoles based on acoustic metamaterials and investigated the associated topological boundary states. By utilizing the coupling parameter as the fourth momentum vector, we reveal two novel topological phases in 3D subsystems of TMSMs: massless Hopf semimetals for the gapless subsystem and chiral topological insulators for the gapped subsystem. We observed the key signatures of both subsystems, the Fermi-arc surface states and Dirac-cone surface states, respectively. Our work provides a practical material platform for realizing topological structures hosting tensor gauge field and opens the door to explore higher-dimensional topological phenomena.

We acknowledge the support of New Cornerstone Science Foundation, the Research Grants Council of Hong Kong (STG3/E-704/23-N, AoE/P-502/20, 17309021), the Research Grants Council of Hong Kong (UGC/FDS24/E04/21), the National Natural Science Foundation of China (12274031), Beijing Institute of Technology Research Fund Program for Teli Young Fellows, National Natural Science Foundation of China (92263208).

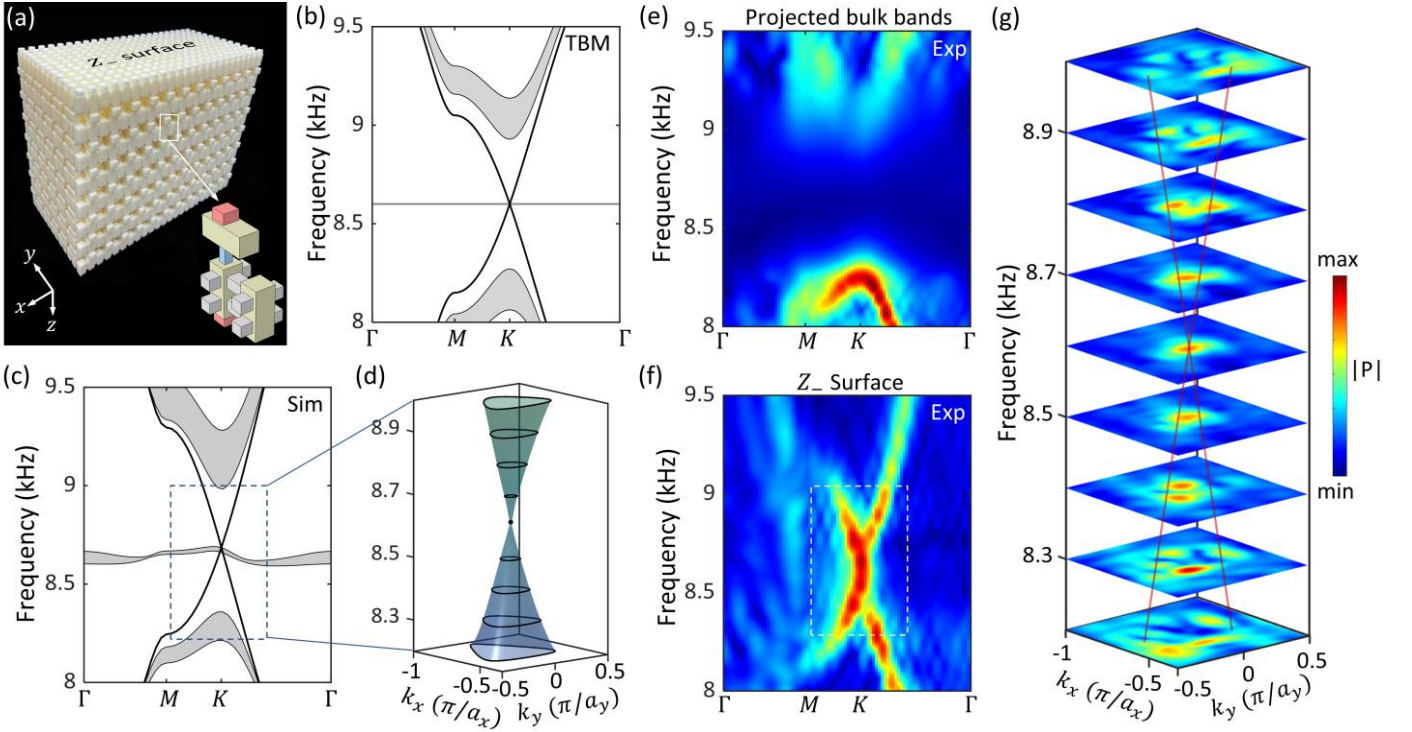

FIG. 4. (a) The photo of acoustic sample for the gapped subsystem, containing $12 \times 12 \times 8$ cells. Inset: the hollow acoustic unit cell with $s_1 = 0.4s_0$ and $s_2 = 1.6s_0$. (b, c) theoretic (b) and simulated (c) projected band spectrum along the $\Gamma - M - K - \Gamma$ high-symmetry line. Black lines and gray areas denote Dirac-cone surface states and projected bulk bands. (d) The zoom-in view of the Dirac-cone surface states in (c). (e, f) Measured spectrum for projected bulk bands (e) and Dirac-cone surface states (f). (g) Measured iso-frequency surfaces for the surface Dirac cone at the $K$ valley.

* These authors contributed equally to this work.
§cuicuilu@bit.edu.cn
‡jiezhu@tongji.edu.cn
†shuzhang@hku.hk